\documentclass[10pt,journal,compsoc]{IEEEtran}

\ifCLASSOPTIONcompsoc
  \usepackage[nocompress]{cite}
\else
  \usepackage{cite}
\fi
\ifCLASSINFOpdf
\else
\fi

\usepackage{algpseudocode}
\usepackage{amsmath,amsfonts}
\usepackage{amsmath,amssymb,amsfonts}

\usepackage{array}

\usepackage{amsmath,amsfonts}
\usepackage{amsmath,amssymb,amsfonts}
\usepackage{algpseudocode}
\usepackage{algorithm}
\usepackage{array}
\usepackage{graphicx}
\usepackage{caption}

\usepackage{url}
\usepackage{fancyhdr}
\usepackage{kantlipsum}
\usepackage{eso-pic}

\begin{document}

\AddToShipoutPictureBG*{%
	\AtPageLowerLeft{%
		\setlength\unitlength{1in}%
		\hspace*{\dimexpr0.5\paperwidth\relax}
		\makebox(0,0.63)[c]{This work has been submitted to the IEEE for possible publication.}
		\makebox(0,0.3)[c]{Copyright may be transferred without notice, after which this version may no longer be accessible.}
}}

%
\title{High-Security Hardware Module with PUF and Hybrid Cryptography for Data Security}
%
%
%
%

\author{Joshua~Tito~Amael,~\IEEEmembership{Member,~IEEE,}
        Oskar~Natan,~\IEEEmembership{Member,~IEEE,}
    and~Jazi~Eko~Istiyanto,~\IEEEmembership{Member,~IEEE}
\IEEEcompsocitemizethanks{\IEEEcompsocthanksitem Joshua Tito Amael is with the Department of Computer Science and Electronics, Universitas Gadjah Mada, Yogyakarta 55281, Indonesia (e-mail: joshua.tito.amael@mail.ugm.ac.id). 
\IEEEcompsocthanksitem Oskar Natan is with the Department of Computer Science and Electronics, Universitas Gadjah Mada, Yogyakarta 55281, Indonesia (e-mail: oskarnatan@ugm.ac.id).
\IEEEcompsocthanksitem Jazi Eko Istiyanto is with the Department of Computer Science and Electronics, Universitas Gadjah Mada, Yogyakarta 55281, Indonesia (e-mail: jazi@ugm.ac.id).}
\thanks{Manuscript received xx xx, xxxx; revised xx xx, xxxx.\\
Corresponding Author: Oskar Natan}}

%
%

\markboth{IEEE Transactions on Computers,~Vol.~xx, No.~xx, xxxx~xxxx}%
{Shell \MakeLowercase{\textit{et al.}}: Bare Advanced Demo of IEEEtran.cls for IEEE Computer Society Journals}
%



\IEEEtitleabstractindextext{%
\begin{abstract}
This research highlights the rapid development of technology in the industry, particularly Industry 4.0, supported by fundamental technologies such as the Internet of Things (IoT), cloud computing, big data, and data analysis. Despite providing efficiency, these developments also bring negative impacts, such as increased cyber-attacks, especially in manufacturing. One standard attack in the industry is the man-in-the-middle (MITM) attack, which can have severe consequences for the physical data transfer, particularly on the integrity of sensor and actuator data in industrial machines. This research proposes a solution by developing a hardware security module (HSM) using a field-programmable gate array (FPGA) with physical unclonable function (PUF) authentication and a hybrid encryption data security system. Experimental results show that this research improves some criteria in industrial cybersecurity, ensuring critical data security from cyber-attacks in industrial machines.
\end{abstract}

\begin{IEEEkeywords}
Cybersecurity, Hardware Security Module, Cryptographic Algorithm, Man-in-the-middle Cyber-attacks
\end{IEEEkeywords}}

\maketitle

\IEEEdisplaynontitleabstractindextext

%
\IEEEpeerreviewmaketitle

\ifCLASSOPTIONcompsoc
\IEEEraisesectionheading{\section{Introduction}\label{sec:introduction}}
\else
\section{Introduction}
\label{sec:introduction}
\fi

\IEEEPARstart{T}{he} development of technology in the industrial world is progressing rapidly. In their research, Frank et al. analyzed that in Industry 4.0, there are already four fundamental technologies: the Internet of Things (IoT), Cloud Computing, Big Data, and Data Analysis \cite{regla_performance_2022}. Additionally, the current industrial progress is also supported by various technologies such as robotics, virtual reality, and cyber-physical systems \cite{tran_transition_2023}. These technologies exist since industries have now shifted to using machines to complete tasks, which are considered more efficient in decision-making and adapting without human intervention . This situation has led many industries to prefer the concept of high automation, relying on automated machines and employing fewer workers \cite{dinlersoz_automation_2023}. However, it also has negative consequences, including an increase in cyber-attacks.

The prevalence of cyber-attacks indirectly contributes to the increase in both quantity and variety of cyber-attacks \cite{amael_securing_2024}. One of the prevalent industrial cyber-attacks is the Man-in-the-middle (MITM) attack. MITM is a cyber-attack where the perpetrator disguises themselves to access the industrial system connecting two users. The perpetrator can monitor or alter sensitive industrial information without detection and may even insert malware viruses that could damage the system. MITM attack schemes in industries typically occur during data or goods transfer processes from manufacturers to consumers. This scheme is based on the common practice of industries conducting physical data transfers, whether for firmware updates, software updates, or other purposes on industrial machines \cite{mallik_man---middle-attack_2019}. Thus, a task-specified module is needed to ensure the security.

\begin{figure} [!t] 
    \centering
    \includegraphics[width = 8.5 cm]{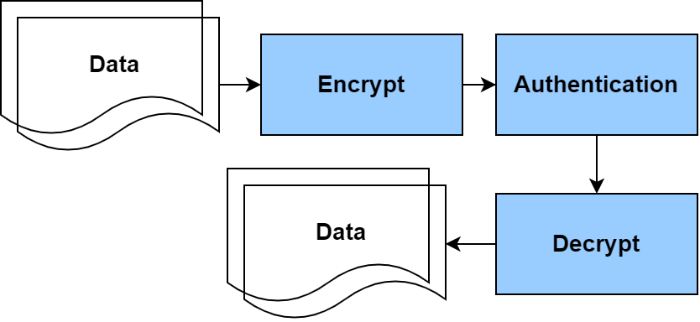}
 \caption{System Overview}
    \label{fig:overview}
\end{figure}

One of the solutions is to employ a hardware security module (HSM), a specialized device capable of performing cryptography operations. HSM is characterized by its ability to manage encryption keys, cryptography operations, and secure information storage, making it difficult to extract \cite{mulder_trends_2023}. In this research, HSM is developed using field programmable gate array (FPGA) microprocessors, incorporating the physical unclonable function (PUF) property for authentication processes, and equipped with a hybrid encryption data security system to ensure data integrity. Protection is specifically dedicated to the cryptographic private key data, which is a crucial part that needs to be safeguarded \cite{hu_styx_2021}. Additionally, this study involves a single-board computer that performs data decryption upon successful authentication processes. The use of FPGA allows for the creation of unique PUF labels, enhancing the distinctiveness and security of the HSM. FPGA is chosen in this research due to its proven robustness in handling various cyber-attacks \cite{benhani_security_2019}. The primary goal of this research is to protect the data as shown in Fig. \ref{fig:overview}. Then, the novelties are listed as follows:

\begin{itemize} 
\item We design a PUF integrated with a hybrid encryption algorithm to enhance the security level of HSM.
\item We innovatively utilize PUF as an authentication method in HSM rather than use it as a cryptography key generation which is very common in most cases.
\end{itemize}

The rest of this paper is organized as follows.  In Section \ref{relatedworks}, we do a comprehensive review of some related works that also inspire this research. In Section \ref{methods}, we explain our proposed model, especially HSM and system evaluation. Then, we analyze the result to understand HSM's stability, data integrity, and time consumption in Section \ref{results}. Finally, we conclude our findings in Section \ref{conclusions}, along with some suggestions for future studies.

\section{Related Works} \label{relatedworks}
This section reviews related works focusing on hardware security modules, cryptography, key management systems, and physical unclonable functions (PUF). Then, we point out the key ideas that inspire our work and serve as an objective for comparative study.

\subsection{Hardware Security Module}\label{AA}
According to Mulder et al., a Hardware Security Module (HSM) is a specialized device capable of cryptography operations to generate and store pairs of private-public keys and secret information related to these keys. HSM is characterized by its ability to generate and manage encryption keys, perform encryption and decryption operations, and securely store information, making it extremely difficult to extract. Some HSMs are also equipped with physical protection, such as tamper-proof features that issue warnings upon interference detection, rendering them inoperable \cite{mulder_trends_2023}.  

The currently developed HSM come in various forms and operational systems. One research on HSM, conducted by Hupp W, focuses on enhancing HSM to secure communication between industrial devices and a server-based system called Module-OT. The HSM development involves securing data using the AES cryptography algorithm and an IP authentication system that blocks foreign IP data connected to the server. This particular HSM still requires third-party testing to assess its security level \cite{hupp_module-ot_2020}

HSM is also developed by Rady et al., who focus on HSM development based on Memristors for securing IoT devices. Memristors are chosen as the main component due to their unique and unpredictable characteristics, making them function as Physical Unclonable Functions (PUF), meaning they cannot be replicated. Each Memristor has a unique behaviour, which will be developed into an AES key for IoT security. This research's limitation lies in using keys limited to AES 128, and the keys generated are only 3 bits in size \cite{rady_memristor-based_2019}. 

The research by Jingwei Hu et al. discusses the practical performance of rank-code-based cryptographic schemes on FPGA platforms with a case study on the quantum-safe Key Encapsulation Mechanism (KEM) scheme based on LRPC codes called ROLLO. ROLLO was one of the candidates in the second round of the NIST post-quantum cryptography standardization process. The presented implementation includes the encapsulation and decapsulation operations of the ROLLO KEM scheme with some variations from the original specifications. The design is fully parameterized, using code-generation scripts to support a wide range of parameter choices for the security levels specified in ROLLO \cite{hu_engineering_2023}.

Jafarzadeh also researched testing FPGAS-based HSMs subjected to malware in the form of Hardware Trojans. The HSM testing involved comparing power variables, delays, and designs before and after the Hardware Trojan attack. The results showed that the Hardware Trojan had minimal impact on power and delays but caused design changes in the FPGAS \cite{jafarzadeh_real_2020}. 

Pott et al. researched HSM, focusing on innovations in data protection for autonomous vehicle industries. The proposed protection system involves using a firmware security module (FSM). FSM is developed with high computation and large memory based on the Infineon Aurix TC399XP microcontroller. FSM is developed using cryptography algorithms in its architecture, specifically AES 128, SHA 256, and ECC 256 algorithms. The research showed that FSM processes were 1.6-4.5 times slower than HSM processing times. However, FSM demonstrated higher flexibility compared to HSM  \cite{pott_firmware_2021}. 

Bathalapalli et al. developed an HSM using the physical uncloneable function (PUF) combined with a trusted platform module (TPM). The developed system first generates keys from the PUF property. After encryption, these keys are uploaded to the server and stored in the TPM storage. The keys can be accessed normally after verification and decryption  \cite{bathalapalli_itpm_2023}. Given the diverse forms of HSMs, this research focuses on the unique properties of PUFs to be applied in FPGASs as an authentication scheme.

\subsection{Cryptography}
The security system implemented in HSM consists of two main components: Key Management and cryptography operations, which include encryption and decryption processes  \cite{qadir_review_2019}. The fundamental concept of cryptography is to encode information or data to achieve information confidentiality, preventing unauthorized parties from deciphering it \cite{wang_file_2011}. The result of cryptography is that data becomes less recognizable to external parties. Research on HSM utilizing cryptography is becoming increasingly prevalent, such as the study conducted by Yilmaz and Ozdemir, which tested IoT devices with various cryptography algorithms, including AES, Blowfish, DES, SHA, ECC, and RSA. The tested devices included measuring the energy consumption performance of IoT devices in terms of MiB/Second and Operation/Second to gauge the number of mebibytes generated by IoT devices in one second of processing \cite{yilmaz_performance_2018}. Additionally, Goswami also conducted similar research, further contributing to the understanding and development of cryptographic implementations in IoT devices \cite{goswami_comparison_2023}. Furthermore, Abdaoui et al. also conducted research to protect IoT systems using fuzzy elliptic curve cryptographic methods, adding a new dimension to the protection and security efficiency of IoT systems \cite{abboudi_design_2022}.

Research was also conducted by Jaspin et al., presenting data protection in the cloud by combining AES and RSA algorithms to form a hybrid encryption system. This amalgamation of algorithms is based on preliminary testing, which demonstrated that AES and RSA algorithms exhibit the highest level of security, fastest execution time, best data integrity, and they produce ciphertext of the same length as the plaintext compared to other algorithms. This research also indicates that a dual-layered cryptography system utilizing AES and RSA algorithms ensures high-security levels, fast execution time, and resilience to error propagation \cite{jaspin_efficient_2021}. 

Prawira et al. researched securing SMS messages using an Arduino microcontroller equipped with an AES cryptography algorithm modified with a chaotic logistic-based PRBG algorithm. The PRGB algorithm is used for key generation in AES message encryption. The encryption system modification using the PRBG algorithm involves adding a random bit sequence as input capable of generating random bits. Additionally, the built system incorporates the SHA algorithm to ensure data integrity, involving preprocessing, padding, and round computation to generate a consistent hash value. This research shows that performance testing of sending encrypted messages ten times resulted in an average transmission time of 2.118 seconds. Meanwhile, sending messages without encryption only required an average time of 1.539 seconds \cite{prawira_p_secure_2020} \cite{grozov_efficiency_2018}. 

Homma et al. introduced a formal method for designing cryptographic processor datapaths based on arithmetic circuits over Galois fields (GF). This method represents GF arithmetic circuits in the form of hierarchical graph structures, where nodes represent sub-circuits whose functions are defined by arithmetic formulas over GFs, and edges represent data dependency between nodes. This approach allows for formal verification through symbolic computation techniques based on polynomial reduction and Gröbner bases. The method's capabilities are demonstrated through an experimental design of a 128-bit AES datapath, including multiplicative inversion circuits over composite fields \cite{homma_toward_2014}.

Altawy et al. discuss minimal design in cryptography for applications with resource constraints. This design focuses on efficient use of hardware area for various cryptographic functions, such as encryption, hashing, authentication, and random bit generation. The authors introduce the sLiSCP family of permutations, which employs two highly efficient cryptographic structures: the 4-subblock Type-2 Generalized Feistel-like Structure (GFS) and a round-reduced, unkeyed version of the Simeck encryption algorithm \cite{altawy_towards_2018}.

Jaiswal and Lata researched message encryption processes using the ECC algorithm on a Xilinx Artix 7 FPGAS microprocessor. This study implemented the ECC algorithm on FPGAs using Montgomery point multiplication, where the message would be transformed into affine points on an elliptic curve device. Message encryption involves several main stages, including key generation, encryption, and decryption. The encryption process involves transforming the message to be encrypted into points on an elliptic curve using the Koblitz method. The results of this research include an analysis of the FPGAS system, including LUTs, LUTRAM, power consumption, and throughput analysis. This research demonstrates that the ECC cryptosystem exhibits high operation frequency and low latency. Additionally, using the ECC algorithm is considered capable of making it difficult for attackers to compromise the security of private keys \cite{jaiswal_hardware_2018}. Furthermore, many studies have implemented cryptographic algorithms in FPGA, particularly for SHA, ECDSA, and Hash Function algorithms \cite{docherty_flexible_2011} \cite{wajih_secure_2008} \cite{genc_design_2021}. In this study, a hybrid encryption system was developed and applied to a single-board computer to secure sensor and actuator data in industrial machines. The selection of this algorithm was based on previous research showing that the combination of AES and RSA algorithms yields high-security levels, fast data processing, and guaranteed data integrity.

\subsection{Key Management System}
In addition to using cryptography algorithms, this research also adheres to the concept of a key management system. The key management system concept is implemented to authenticate key storage within the FPGAS. The basis for choosing this concept comes from the research of Luo et al. This study manages keys at the cloud level using an ARM board. It is based on centralized cloud services that cannot share the same keys with others. Therefore, in their research, TZ-KMS becomes a solution for secure key management. The security system uses a centralized KMS architecture to access cloud servers. The connection between the cloud is secured using ARM TrustZone, and to ensure hardware security, a system-on-chip (SoC) is used. TZ-KMS is divided into several services: user identification, key management (creating, deleting, and managing keys), and cryptography operations. This research shows that TZ-KMS can withstand DoS attacks, but it has yet to be resilient against large-scale DoS attacks. Additionally, Luo et al. ensure that large-scale DoS attacks will not compromise the data but may cause slight performance overheating \cite{luo_tz-kms_2018}.

shim's research designs a quantum key management system for QKD, suitable for KREONET. It manages symmetric key lifecycles, supports many-to-many and P2P communication, and addresses QKD distance limitations validated via simulations \cite{shim_design_2022}

Xia's paper proposes a synthetic key management system to enhance the security and efficiency of power distribution operations by combining quantum private communication and automatic key management. It analyzes the feasibility of this integration, designs a key-management method involving encryption devices, and explores a rollback mechanism for large-capacity quantum key storage to ensure integrity and atomicity \cite{xia_research_2019}.

This research is also inspired by the study by Gu et al., which introduced the Key Tree Reuse (KTR) scheme in a wireless key management system. This system emphasizes data access control and is considered effective for distribution to many users. In this study, efforts are made to secure private keys and authenticate multiple data wirelessly \cite{gu_ktr_2009}.

\subsection{Physical Uncloneable Function}
Physical Unclonable Function (PUF) is a classical security technology that extracts bits originating from distinctive variations in each Integrated Circuit (IC) production, where each hardware device has unique, unpredictable characteristics that prevent cloning. PUF can also be interpreted as a function derived from a physical system that cannot be cloned. The system responds with a unique response when a stimulus is applied \cite{balan_puf-based_2020} \cite{shamsoshoara_survey_2020}.  According to Vishal Pal et al., PUF is akin to a fingerprint, ensuring that each instance is always unique from another \cite{pal_puf_2020}.

The PUF has a structure that generates a response (R), the only output to the input called challenge (C). Therefore, it can be modeled as a function representing the input-output relationship. The device is the input set challenge C={C1, C2,…, Cn} for the set of outputs response R={R1, R2,…, Rn}=PUF (C1, C2,…, Cn) that can be extracted from PUF, and all challenge-response pairs, CRPs={(C1, R1),(C2, R2),…,(Cn, Rn)} can be obtained. Secret key generation and device authentication are performed using this CPR information. PUF can be divided into two categories: weak PUF and strong PUF. Weak PUF typically has no fixed challenge and has a limited response. Therefore, it is mainly used for secret key or identity generation. Strong PUF can acquire many CRPs. Since it is difficult to predict the response to a random challenge, it is often used for device authentication \cite{kim_reinforcement_2022}.

Research on PUF is also conducted by Kumar et al., analyzing delay-based PUF categorization on FPGAS and the potential of PUF in security and authentication systems. In this study, testing is conducted by providing 8-bit challenge inputs with ten variations. The results of this research testing indicate that the input-output pairs of PUF determine whether the PUF is strong or weak. PUFs have a very large and exponentially growing CRP, allowing users to validate response conventions where each challenge-response space is used only once. Meanwhile, weak PUFs have limited CRP concerning their application size. Weak PUFs are used for secure key storage \cite{kumar_fpga_2017}. 
Kim et al., in their research, categorized the strengths of various types of PUFs as shown in Table 1 \cite{kim_secure_2020}.

\begin{table}
    \centering
        \caption{PUF Comparison}
    \label{tab: PUF Comparison}
    \begin{tabular}{|>{\raggedright\arraybackslash}p{2cm}|>{\raggedright\arraybackslash}p{2cm}|>{\raggedright\arraybackslash}p{2cm}|} \hline 
         PUF Type&  Advantages& Disadvantages\\ \hline 
         Arbiter PUF&  Robust against machine learning attacks& Delay pass must be the same\\ \hline 
         RO PUF&  Easy to implement& Sensitive to environmental factors\\ \hline 
         SRAM PUF&  Good statistical performance& Limited number of CRPs
\\ \hline 
         VIA PUF&  No ECC Module& Limited number of CRPs
\\ \hline 
         PDRO PUF&  Have more CRPs than RO-PUF& Sensitive to environmental factors
\\ \hline 
         PHY PUF&  Real-time acquisition& Limited number of CRPs
\\ \hline
    \end{tabular}

\end{table}

The PUF system we developed is also inspired by the system developed by Oun and Niamat., which designed a Ring Oscillator PUF for authentication in IoT device security. The authentication process consists of two types: the enrollment process, which is connected to the server, and the verification process, which compares the authentication response results. This study used 10 Artix-7 FPGAs and demonstrated that PUF suits hardware security authentication schemes \cite{oun_puf-based_2023}.

Hemavathy and Bhaaskaran researched the security aspects of PUF, especially the Arbiter PUF. The results of this study indicate that the Arbiter PUF is a strong security-wise, delay-based PUF depending on the manufacturing characteristics of the hardware. Additionally, it is demonstrated that the CRP possessed by the Arbiter PUF has an exponential number, making it suitable for authentication with low resource utilization \cite{hemavathy_arbiter_2023}. Therefore, based on previous research, this study implements the development of an arbiter PUF used for the authentication scheme of private keys in a cryptography algorithm within an HSM.

\section{Method} \label{methods}
This section describes the architecture of the hardware security module (HSM), hybrid encryption schema, and physical unclonable function authenticator. Then, we also conduct system evaluations to ensure the system runs well regarding stability, security, and data integrity.

\subsection{Proposed Model}\label{AA}
\begin{figure*} [t] 
    \centering
    \includegraphics[width = 17 cm]{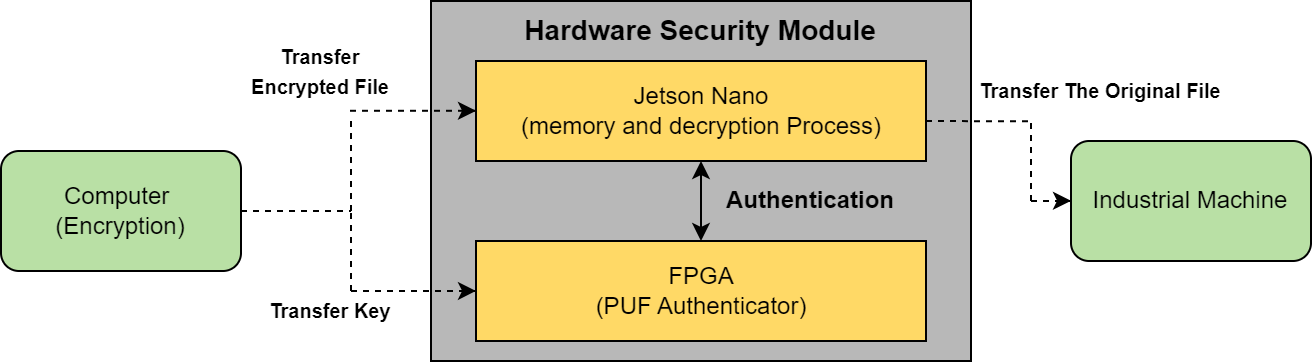}
    \caption{Hardware Security Module Architecture}
    \label{fig:HSM_Arch}
\end{figure*}

The hardware security module is implemented into two main parts: the hybrid encryption scheme and the physical unclonable function authenticator. The HSM system involves several key devices, including an FPGAS, a Single Board Computer (SBC) such as the Jetson Nano, and a computer. Each of these devices has its function, with the computer serving as the location for the hybrid encryption system, the FPGAS authenticator acting as the medium for the authentication process using PUF, and the SBC serving as the storage for encrypted files and conducting the data decryption process when authentication is successful. The full-scale implementation of the hardware security module is depicted in the architecture shown in Fig. \ref{fig:HSM_Arch}.

\begin{figure*} [t!] 
    \centering
    \includegraphics[width = 17 cm]{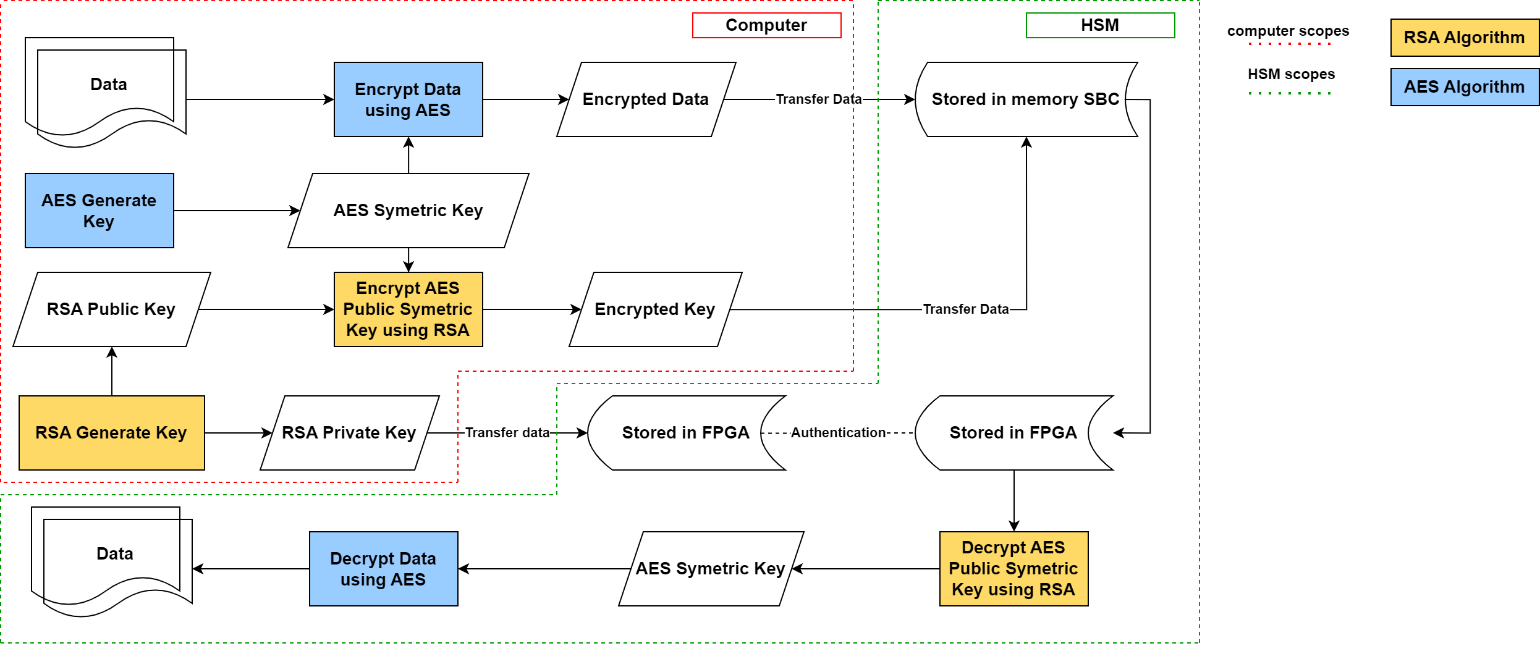}
    \caption{Technical Diagram of HSM System}
    \label{fig:TD_HSM}
\end{figure*}

\begin{figure*} [t!] 
    \centering
    \includegraphics[width = 17 cm]{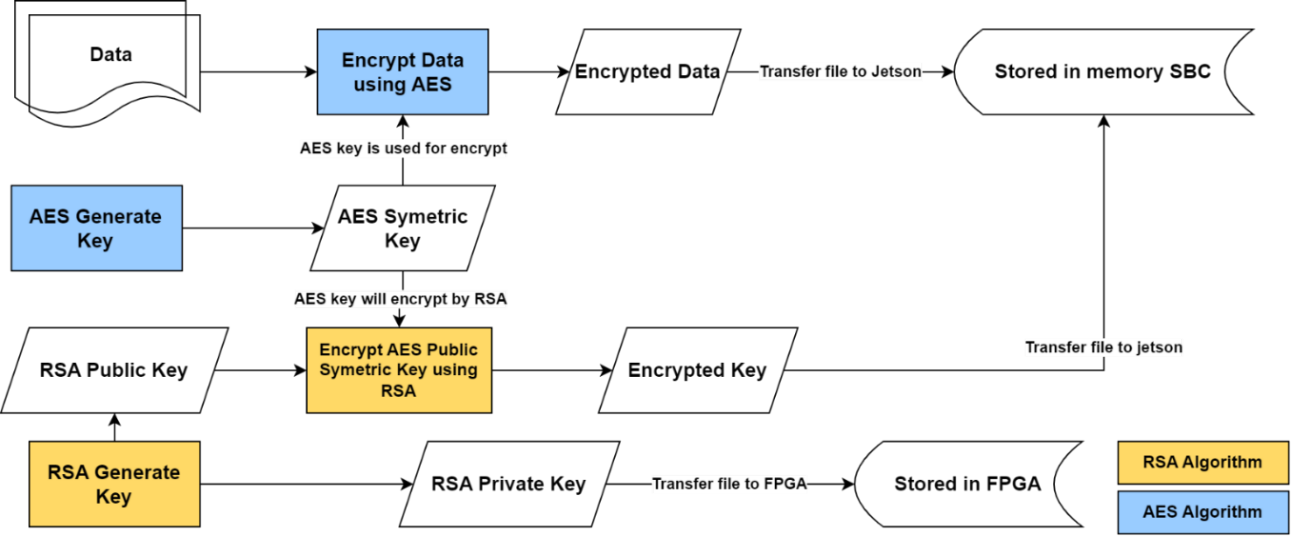}
    \caption{Hybrid Encryption Schemes}
    \label{fig:Enc_scheme}
\end{figure*}

The HSM system designed in this study will begin by encrypting data on the computer using a dual encryption system. This dual encryption process generates encrypted data files and a private key used for decrypting the files. The encrypted data files will then be transferred and stored in the memory of the Jetson Nano SBC. Meanwhile, the private key will be transferred to the FPGAS board for the authentication key process. This authentication process is intended to verify whether the key being stored is the key to unlocking the previously encrypted files. When this authentication process succeeds, the FPGAS will send a signal via UART to the Jetson Nano. The decryption process is initiated when the user requests decryption. When the user requests to proceed with the data decryption process, the decryption will be carried out within the Jetson Nano SBC, and the file will be ready for reuse and transferred to the industrial machine. The technical details of the system within the HSM are illustrated in Fig. \ref{fig:TD_HSM} and \ref{fig:Enc_scheme}.

The technical HSM system begins with generating an AES algorithm key for decryption purposes. The data to be protected is then encrypted within the computer using the AES algorithm. The process proceeds with storing the encrypted files in the Jetson Nano memory. As for the symmetric AES key previously used for encryption will be encrypted again using the RSA algorithm, which naturally involves generating public and private keys beforehand. The result of this encryption will be transferred to the Jetson Nano in the form of an encrypted AES key. The RSA algorithm is an asymmetric cryptography algorithm requiring a private key for decrypting the AES key. Therefore, the RSA's private key in this system will be sent to the FPGAS and entered into the PUF circuit. The FPGAS will authenticate the response from the key entering the PUF. If the response value from the key has yet to be registered in the FPGAS, the authentication process is deemed unsuccessful, and decryption cannot proceed on the Jetson Nano.
Using the private RSA key, the decryption continues to the next step. Conversely, if the response value matches the registered value in the FPGAS, the authentication process is deemed successful, and decryption proceeds. First, decryption is performed on the AES key, followed by decryption of the data to be secured. Upon completion of the process, the data will be ready for use in industrial machines. In broad terms, this HSM system is divided into two main parts: the hybrid encryption scheme and the physical unclonable function authenticator.

\subsection{Hybrid Encryption Scema}
The hybrid encryption system combines two algorithms, AES and RSA, selected based on their combined capabilities in execution speed, data integrity, and high security. The selection of these algorithms is grounded on the research by Jaspin et al in Tabel II \cite{jaspin_efficient_2021}. The system begins by encrypting data using the AES algorithm, as shown in Algorithm 1. In its application, the AES algorithm undergoes looping to obtain random values and utilizes encoding principles through iteration. Subsequently, the encrypted data will be stored in the Jetson Nano memory. The system proceeds with the second encryption process, which involves encrypting the symmetric AES key using the RSA algorithm. This encryption process uses the RSA concept outlined in Algorithm 1.

\begin{table*}
    \centering
        \caption{Comparison of Proposed Work With Existing Method \cite{jaspin_efficient_2021}}
    \label{tab:my_label}
    \begin{tabular}{|c|c|c|c|c|c|} \hline 
         Parameters&  DES&  Blowfish&  RC5&  3 DES& AES + RSA\\ \hline 
         Security&  Not Secure&  Secure&  Partially Secure&  Better than DES& Very Secure\\ \hline 
         Speed&  Very Slow&  Fast&  Slow&  Slow& Very Fast\\ \hline 
         Data Confidentiality&  No&  Yes&  No&  No& Yes\\ \hline 
         Data Integrity&  No &  Yes&  No&  No& Yes\\ \hline 
         Cipher Text&  Larger than plaintext&  Same as Plaintext&  Larger than plaintext&  Larger than plaintext& Same as plaintext\\ \hline
    \end{tabular}

\end{table*}

\begin{algorithm}
\caption{AES Encryption Algorithm}
\begin{algorithmic}[1]
    \Require Plaintext $P$, Key $K$
    \Ensure Ciphertext $C$
    
    \State \textbf{KeyExpansion:} Expand the key $K$ to generate round keys.
    
    \State \textbf{InitialRound:}
    \State AddRoundKey($P$, round\_key[0])
    
    \For{$i = 1$ to $N-1$} \Comment{Repeat for $N-1$ rounds}
        \State \textbf{SubBytes:} Substitute each byte of the state with the corresponding byte from the S-box.
        \State \textbf{ShiftRows:} Rotate the rows of the state.
        \State \textbf{MixColumns:} Transform columns of the state.
        \State \textbf{AddRoundKey:} XOR the state with the round key.
    \EndFor
    
    \State \textbf{FinalRound:}
    \State SubBytes
    \State ShiftRows
    \State AddRoundKey(round\_key[$N$])
    
    \State \textbf{Output:} Ciphertext $C$
\end{algorithmic}
\end{algorithm}

\begin{algorithm}
\caption{RSA Algorithm}
\begin{algorithmic}[2]
    \State $n \gets p \times q$
    \State $\phi(n) \gets (p-1) \times (q-1)$
    \State Find $d$ such that $d \times e \equiv 1 \mod \phi(n)$
    \State $p, q \gets$ random prime numbers
    \State $e \gets$ integer $(1 < e < \phi(n))$
    \State $e, n \gets$ public key (encryption)
    \State $d, n \gets$ private key (decryption)
\end{algorithmic}
\end{algorithm}

\begin{figure} [h] 
    \centering
    \includegraphics[width = 9 cm]{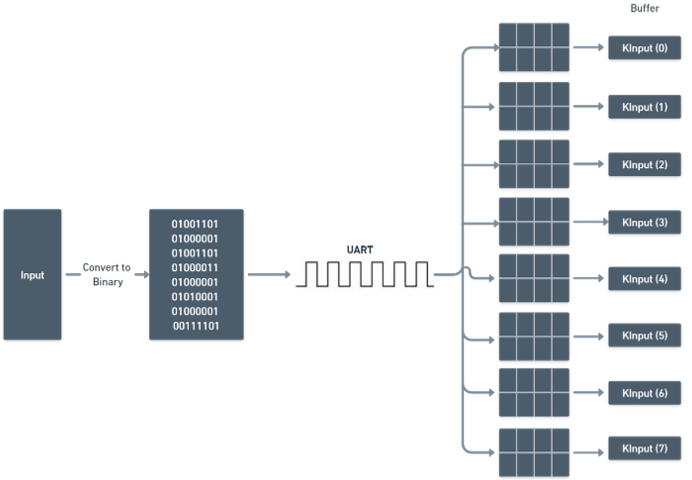}
    \caption{Key Input Process in FPGAs}
    \label{fig:key}
\end{figure}

The process then continues by transferring the encrypted AES key to the Jetson Nano. Meanwhile, the private RSA key will enter the FPGAs system using the UART graphical user interface. In the UART interface, the user will input the private RSA key into the FPGAS and enter a PIN passphrase agreed upon by multiple parties. Subsequently, the key will be entered into the FPGAS and ready for processing. Before the key enters the FPGAS, the GUI will automatically convert the key file into a binary value. Upon entering the FPGAS, this binary value will be divided into eight buffers for processing into the PUF circuitry, as illustrated in Fig. \ref{fig:key}.\newline

\subsection{PUF Authentication}
\begin{figure*} [t!] 
    \centering
    \includegraphics[width = 15 cm]{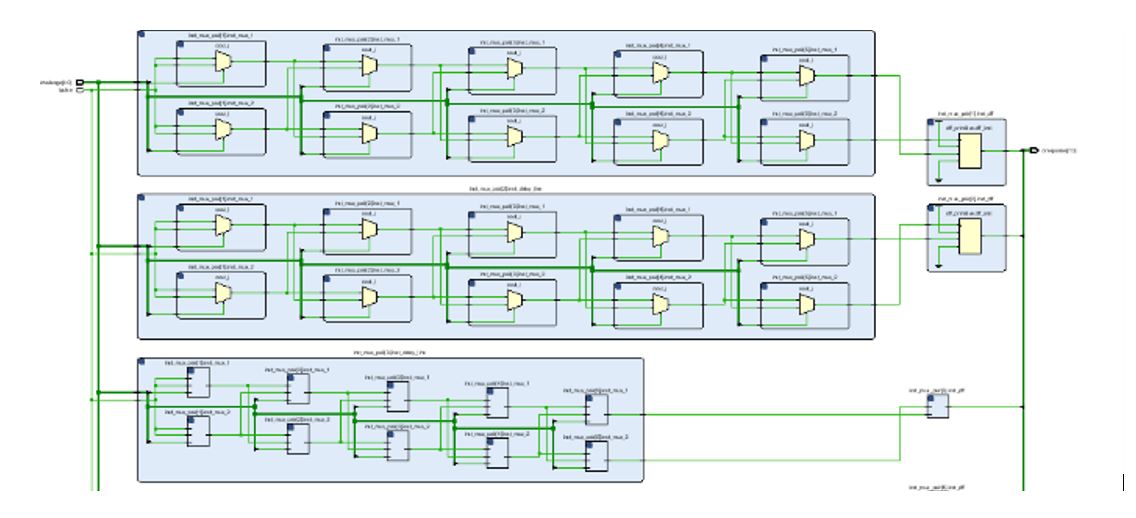}%
    \caption{Arbiter PUF design Architecture}
    \label{fig:arbiter}
\end{figure*}

\begin{figure} [t!] 
    \centering
    \includegraphics[width = 9 cm]{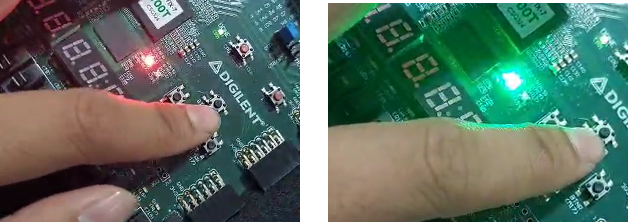}
    \caption{PUF Authentication Process Fail (Left) \& Authentication Sucess (right)}
    \label{fig:auth}
\end{figure}

The Physical Unclonable Function (PUF) is utilized as an authenticator to verify whether the key inputted into the FPGAS is correct or incorrect. This can be likened to facial authentication used when you want to unlock your smartphone \cite{li_empirical_2018}. In its implementation, the FPGAS used in this research is the Artix 7 Nexys 4. The operation of this PUF is illustrated by marking unique codes or characteristics within an object. Subsequently, the response from this PUF will be organized as an authenticator for the response already registered within the FPGAS. The type of PUF utilized in this research is the arbiter PUF, which consists of multiplexer logic gates intertwined with D-Flip-flops, as shows in Fig. \ref{fig:arbiter}.
The PUF will then produce a response in the form of a binary value. Authentication is successful if this binary value is registered in the FPGAS memory. Conversely, the authentication process is only possible if the binary value is found in the FPGAS memory list. A failed authentication process also indicates that the key inserted into the PUF is a foreign key previously unrecognized by the FPGAS system. A green light indicates the success of authentication in the FPGAS, while failure will be indicated by a red light, as shown in Fig. \ref{fig:auth}.

\subsection{System Evaluation}

System testing is conducted to ensure that the HSM system operates with the uniqueness of the PUF, ensuring security and data integrity during processing and the efficiency of the time required for data processing. This is also in line with several studies that have tested the speed of RSA algorithms on FPGA \cite{asokan_performance_2020} \cite{havrilov_hardware_2020}. These 3 tests are conducted to determine whether the HSM can function correctly. The authentication response is tested through multiple trials to determine the uniqueness of the PUF's HSM system.
Data integrity testing is performed by comparing data visualization before and after processing to ensure the integrity of the data and determine whether it is corrupted. The final test assesses how quickly the developed HSM can process data during hybrid encryption. The speed of execution of this HSM is calculated using Htop. This efficiency testing is crucial for an HSM to ensure it meets the required performance standards and can handle encryption processes promptly \cite{baee_provably_2024}.

\section{Result} \label{results}
The results of several HSM tests are as follows:

\subsection{Uniqueness of Hardware Security Module}
The effectiveness testing of the HSM was conducted by performing 22 experiments for the authentication process using different keys. The experiments were divided into two conditions: Experiment 1, which included 11 different types of keys, and Experiment 2, which included 11 keys of the same type as those in Experiment 1. The results of the testing can be seen in Table II.

The results of these experiments show that the PUF of the FPGAs has a uniqueness value of 72\%. This uniqueness value was calculated with 22 input challenge experiments that produced 16 different responses. Therefore, the uniqueness value generated from this PUF is 72\%. The high uniqueness value in the PUF FPGA can be attributed to the inherent manufacturing variations in the FPGA fabric. These variations cause each FPGA to respond differently to the same input challenges, resulting in unique and unpredictable responses. Additionally, the structure and design of the PUF circuit within the FPGA are optimized to enhance these variations, thereby maximizing the uniqueness and reliability of the responses generated.

\begin{table*}
    \centering
        \caption{PUF Design Uniqueness Test Result}
    \begin{tabular}{|c|c|c|c|c|c|} \hline 
         \multicolumn{3}{|c|}{\textbf{Experiment 1}}&  \multicolumn{3}{|c|}{\textbf{Experiment 2}}\\ \hline 
 \textbf{Input Key}& \textbf{Response}& \textbf{Authentication}& \textbf{Input Key}& \textbf{Response}& \textbf{Authentication}
\\ \hline 
         0000000000000000&  1111111111011110&  V&  0000000000000000&  1111111111011110& v
\\ \hline 
         1000000000000000&  1111111111011110&  V&  1000000000000000&  1111111111011110& v
\\ \hline 
         0010000000000000&  1111111111011110&  V&  0010000000000000&  1111111111011110& v
\\ \hline 
         0001000000000000&  1111111111011110&  V&  0001000000000000&  1111111111011110& v
\\ \hline 
         0100000000000000&  0111111111011110&  x&  0100000000000000&  0111111111011110& 
x\\ \hline 
 0000100000000000& 1111111111111110& x& 00001000 00000000& 1111111111111110&
x\\ \hline 
 0000010000000000& 1111111111111010& x& 0000010000000000& 1111111111111010&
x\\ \hline 
 0000001000000000& 1111111111011000& x& 0000001000000000& 1111111111011000&
x\\ \hline 
 0000000100000000& 1111111011111100& x& 0000000100000000& 1111111011111100&
x\\ \hline 
 0000000010000000& 1001111111111101& x& 0000000010000000& 1001111111111101&
x\\ \hline 
         0000000000000001&  1001000010101001&  x&  0000000000000001&  1001000010101001& 
x\\\hline
    \end{tabular}

\end{table*}

\subsection{Data Integrity of Hardware Security Module}
\begin{figure} [t] 
    \centering
    \includegraphics[width = 9 cm]{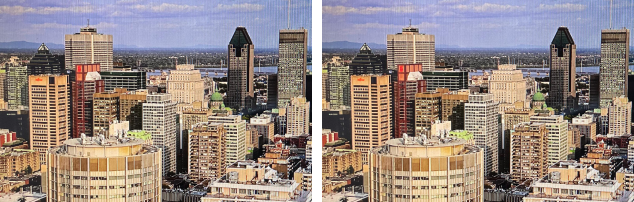}
    \caption{Image Before Processing (left) \& After Processing (right)}
    \label{fig:process}
\end{figure}

One of the most important components in securing HSM is maintaining the integrity and authenticity of the secured data. Therefore, in this test, a comparison was made between image data before and after processing to ensure data integrity. Image data was chosen for this test because it can be visually inspected and is generally easier to compare. Fig. \ref{fig:process} presents the results of the data integrity test on the HSM.

\begin{figure*} [t] 
    \centering
    \includegraphics[width = 18 cm]{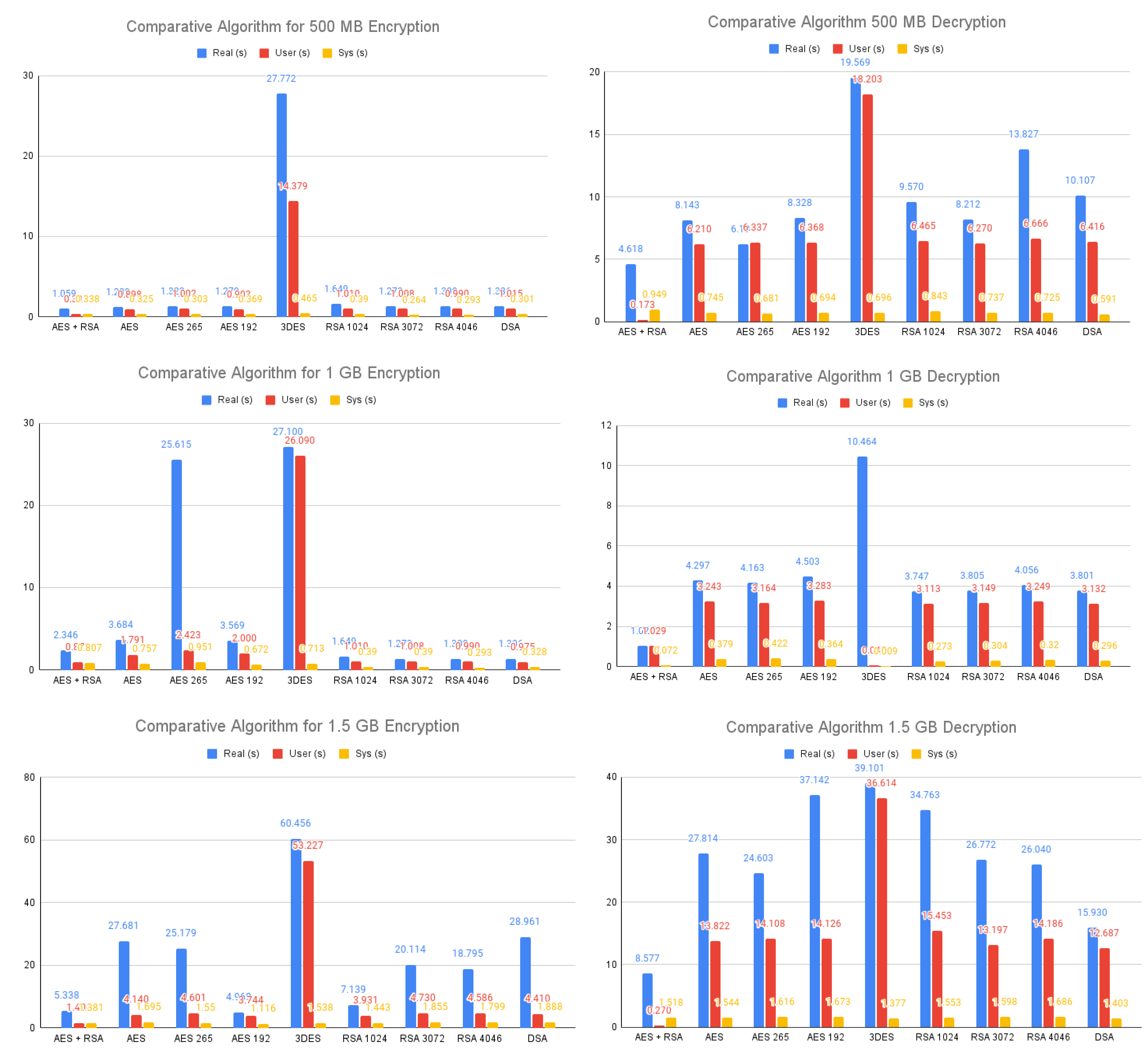}
    \caption{Comparative Algorithm for Cryptographic}
    \label{fig:comparative}
\end{figure*}

The testing proves that the tested image data, with a size of 500.23 MB, maintains the same visual image appearance before and after processing. Additionally, there is no change in the data size before and after processing, remaining at 500.23 MB. This systematic approach guarantees the proper key, ensuring that data is always restored to its original state as shown Fig. \ref{fig:process}. Furthermore, AES and RSA are designed to operate at the bit level of data, executing intricate transformations and enhancing security by spreading the impact of data bits throughout the cipher block, thus minimizing the chance of data corruption. The testing confirms that data handled within the HSM will be protected against processed data corruption. Furthermore, these results represent an improvement from the research conducted by Murali and Prasad on the comparative study of AES and RSA cryptographic algorithms \cite{murali_comparison_2017}

The integrity of the data remains intact within the HSM due to several reasons. Robust encryption mechanisms, such as AES (Advanced Encryption Standard) and RSA (Rivest-Shamir-Adleman), play a crucial role. AES operates at the bit level, ensuring that each bit of the plaintext is transformed and diffused throughout the cipher block, making it exceedingly difficult for any bit-level alteration to occur without detection. RSA provides strong encryption for secure key exchanges, ensuring that the keys used for AES encryption are protected. This layered approach enhances overall security, ensuring that data remains intact during both encryption and decryption processes. The use of PUF (Physical Unclonable Function) for authentication adds an additional layer of security. PUFs leverage inherent manufacturing variations in hardware to generate unique and unclonable responses to challenges, ensuring that only authenticated and legitimate HSM devices can access and process the data, thereby protecting against unauthorized modifications.

Data integrity verification is further ensured by comparing image data before and after processing, confirming that the visual appearance and size remain unchanged. This visual and size consistency check is a straightforward yet powerful method to verify that no data corruption has occurred during processing. By ensuring end-to-end encryption with AES and RSA, the data is protected from the moment it is encrypted until it is decrypted, ensuring that data cannot be tampered with or corrupted at any stage of its lifecycle. Proper key management practices guarantee that the correct keys are used for encryption and decryption, ensuring that data can always be restored to its original state without any integrity loss. These combined factors ensure that the integrity of the data remains preserved within the HSM, providing robust protection against any potential corruption during data processing.

\subsection{Time Consumption of Hardware Security Module}

Time consumption is one of the critical aspects of HSM processing. The time consumed by an HSM should be fast in processing data. This testing evaluates the time required when the HSM processes hybrid encryption. The testing uses the Htop feature on the Linux OS to facilitate time calculations. Ten experiments are conducted for each process and data size and then averaged. Time consumption testing includes three types of time: real-time, user time, and system time. Real-time is the actual time required for processing. User time is the CPU time used during processing. System time is the CPU time processes use in the operating system kernel during processing. Table IV shows the results of the time consumption testing for the developed HSM.

The time consumption testing shows that, on average, the encryption process takes 1.059 seconds with a scale of 10 attempts for a file size of 500 MB. Similarly, the decryption process takes approximately 1.029 seconds. When the file size is increased, the encryption process takes an average of 2.346 seconds, while decryption requires an average of 4.618 seconds. The largest file in the test, 1.5 GB, takes an average of 5.338 seconds for encryption and 8.577 seconds for decryption. The speed of this encryption process is proportional to the increasing amount of data processed inside the HSM. However, this encryption process is still considered fast because the encryption cycle is conducted in a single, comprehensive process that includes subByte, ShiftRows, Mix Columns, and AddRoundKey, executed concisely to shorten the algorithm and enhance processing speed.

Based on Fig. \ref{fig:comparative} proves that the combination of AES and RSA algorithms represents one of the best cryptographic solutions in terms of time consumption when compared to other algorithms like 3DES, DSA, AES (with different key lengths such as AES192 and AES256), and RSA (with varying key sizes such as RSA1048, RSA3072, and RSA4096). This optimal performance is evidenced in the time consumption tests, where the combination of AES and RSA provides robust security and achieves efficient processing times. 

In conclusion, the AES and RSA combination stands out as the best cryptographic solution from a time consumption perspective. It offers a robust security framework while maintaining efficient processing times, making it ideal for securing large amounts of data without compromising on performance.

\begin{table}
    \centering
        \caption{Time Consumpt Test Result}
    \label{tab:Time Consumpt Test Result}
    \begin{tabular}{|c|c|c|c|c|} \hline 
         File Size&  Process&  Real (s)&  User (s)& Sys (s)
\\ \hline 
         500 Mb&  Encrypt&  1,059&  0,337& 0,338
\\ \hline 
         500 Mb&  Decrypt&  1,029&  0,072& 0,445
\\ \hline 
         1 Gb&  Encrypt&  2,346&  0,879& 0,807
\\ \hline 
         1 Gb&  Decrypt&  4,618&  0,173& 0,949
\\ \hline 
         1.5 Gb&  Encrypt&  5,338&  1,406& 1,381
\\ \hline 
         1,5 Gb&  Decrypt&  8,577&  0,270& 1,518
\\ \hline
    \end{tabular}

\end{table}

\section{Conclusions} \label{conclusions}
The design of the hardware security module consists of two main components: hybrid encryption and physical uncloneable function authenticator. From the conducted testing, it was found that the HSM has a high level of PUF stability. This was evidenced by two identical tests producing the same results. The high stability of PUF indicates that the randomness property of PUF can be controlled. The developed HSM also exhibits good data integrity, as shown in Table III. The tested image data remained unchanged, and the resulting data size remained consistent before and after processing, indicating that data processed in the HSM will not be corrupted. One crucial aspect of HSM is time consumption. This study showed that the processing time, especially in the hybrid encryption system, is sufficiently fast. This research is likely further developed regarding features and performance. 
As for future studies, the research can be extended to add a PUF layer for authentication, and hybrid encryption methods can be developed by adding modern cryptography algorithms such as the Elliptic Curve (EC) and Elliptic Curve Digital Signature Algorithm (ECDSA).

\ifCLASSOPTIONcompsoc
  \section*{Acknowledgments}
\else
  \section*{Acknowledgment}
\fi

This work was partially supported by the Department of Computer Science and Electronics, Universitas Gadjah Mada under the publication Funding Year 2024.

\ifCLASSOPTIONcaptionsoff
  \newpage
\fi




%
\bibliographystyle{IEEEtran}
\bibliography{citation}

%

\begin{IEEEbiography}[{\includegraphics[width=1in,height=1.25in,clip,keepaspectratio]{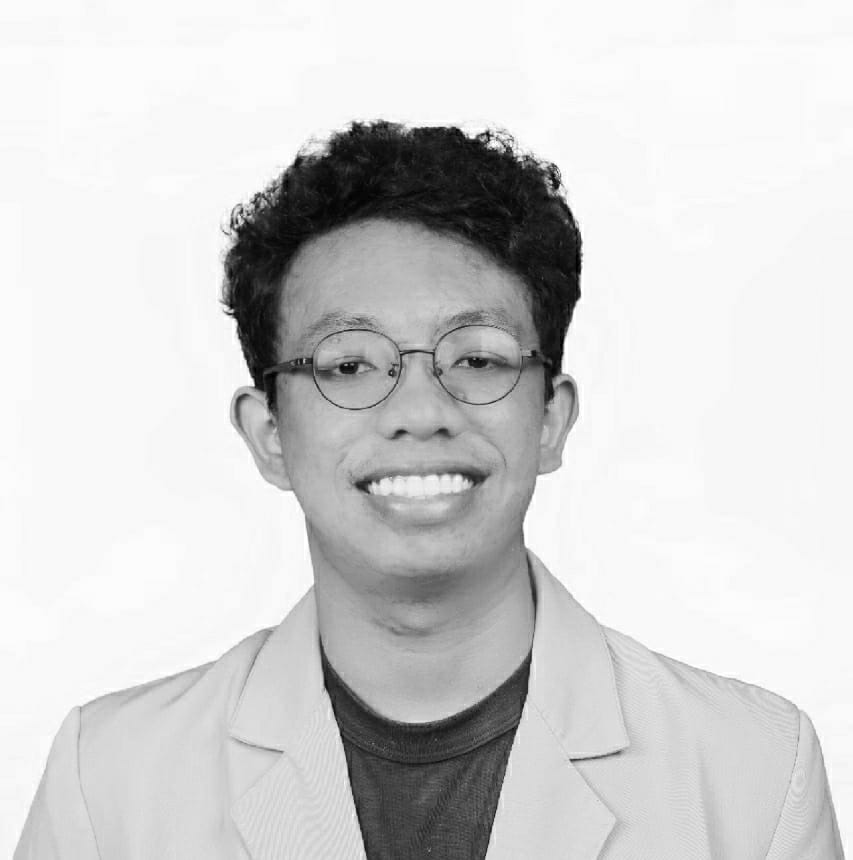}}]{Joshua Tito}
(Member, IEEE) is an undergraduate student at Electronic and Instrumentation, Universitas Gadjah Mada, Indonesia, in 2021. He has been actively involved in research and development in the fields of hardware programming, cybersecurity, and artificial intelligence. In 2024, he was recognized as an awardee best paper for students at the 33rd International Symposium on Industrial Electronics (ISIE) in Ulsan, South Korea. Since February 2024, he has served as an Assistant at the Electronics and Instrumentation Lab at Universitas Gadjah Mada. His research interests include hardware security modules, FPGA, and cryptographic systems.
\end{IEEEbiography}
\vfill
\begin{IEEEbiography}[{\includegraphics[width=1in,height=1.25in,clip,keepaspectratio]{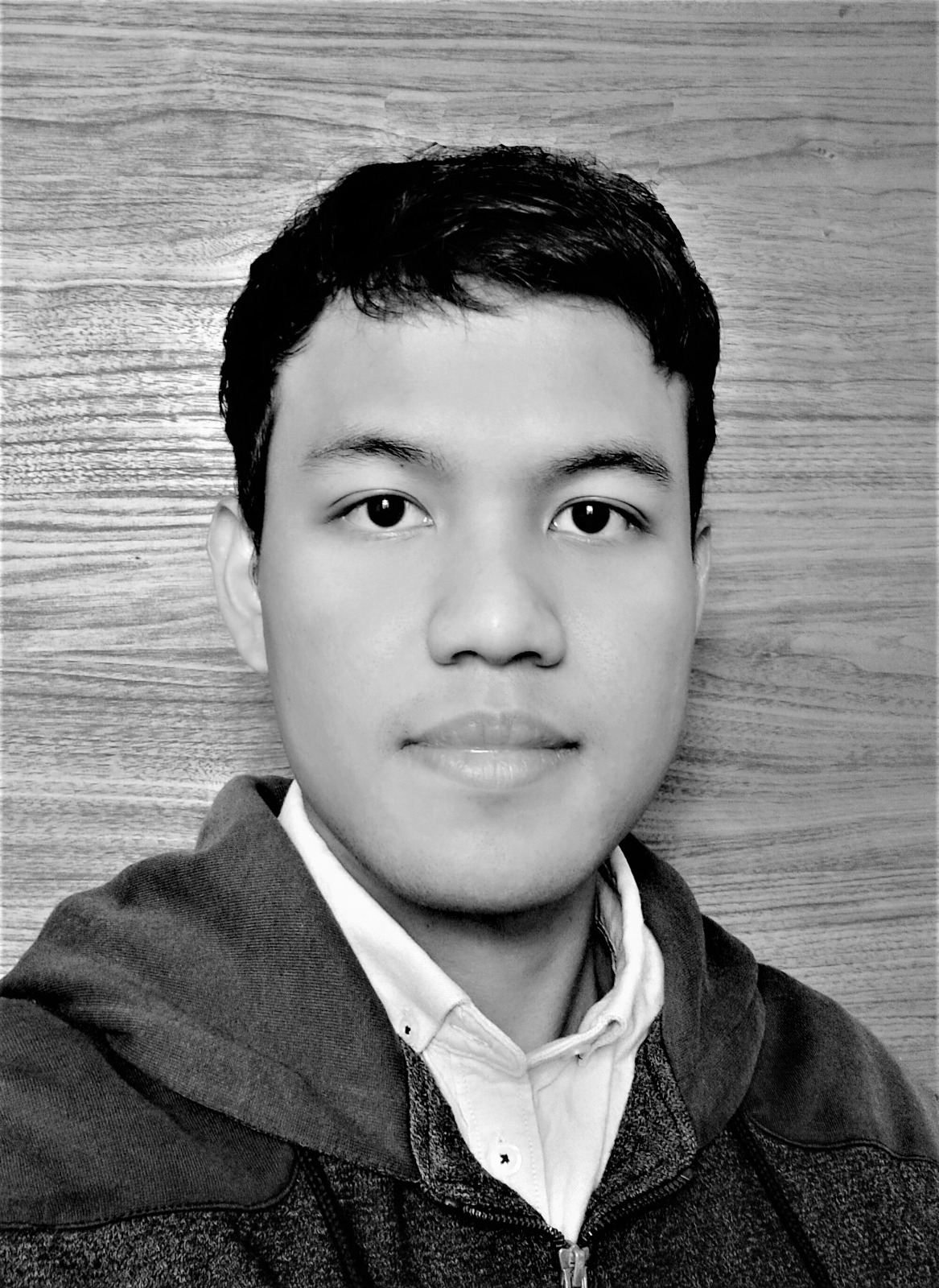}}]{Oskar Natan}
(Member, IEEE) received his B.A.Sc. in Electronics Engineering and M.Eng. in Electrical Engineering from Politeknik Elektronika Negeri Surabaya, Indonesia, in 2017 and 2019, respectively. In 2023, he received his Ph.D.(Eng.) in Computer Science and Engineering from Toyohashi University of Technology, Japan. Since January 2020, he has been affiliated with the Department of Computer Science and Electronics, Universitas Gadjah Mada, Indonesia, first as a Lecturer and currently serves as an Assistant Professor. His research interests lie in the fields of sensor fusion, hardware acceleration, and end-to-end systems for various computer science and electronics applications.
\end{IEEEbiography}

\begin{IEEEbiography}[{\includegraphics[width=1in,height=1.25in,clip,keepaspectratio]{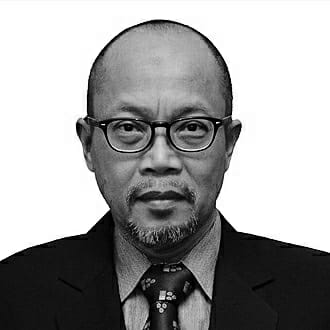}}]{Jazi Eko Istiyanto}
(Member, IEEE) received his B.Sc. in Physics from Universitas Gadjah Mada (UGM), Indonesia, in 1986. Then, he received his M.Sc. in Computer Science and Ph.D. in Electronic Systems Engineering from the University of Essex, UK, in 1988 and 1995, respectively. Since March 1988, he has been serving as a Lecturer at UGM. In August 2010, he became a Professor of Electronics and Instrumentation at the Department of Computer Science and Electronics, Faculty of Mathematics and Natural Sciences, UGM. He was appointed the 5th Head of the Republic of Indonesia's Nuclear Energy Regulatory Agency (BAPETEN) on February 7, 2014. To date, Prof. Jazi has authored and co-authored many papers published in internationally reputable journals and conferences. His research interests include embedded systems, cyber-security, and performance evaluation.

\end{IEEEbiography}
\vfill



\end{document}